\documentclass[preprint,12pt,authoryear]{elsarticle}

\usepackage{amssymb}
\usepackage{comment}
\usepackage{lineno}
\usepackage[utf8]{inputenc}
\usepackage[english]{babel}
\usepackage{subcaption}
\usepackage{minted}
\usepackage{booktabs}
\usepackage{multirow}
\usepackage{hyperref}

\hypersetup{
    colorlinks = true,
    linkcolor= blue,
    citecolor= red,
}

\journal{University of Antwerp}

\begin{document}

\begin{frontmatter}



\title{Design and Implementation of a Domain-specific Language for Modelling Evacuation Scenarios Using Eclipse EMG/GMF Tool}


\author{Heerok Banerjee}
\address{Dept. Of Mathematics \& Computer Science\\ University of Antwerp}
\ead{Heerok.Banerjee@student.uantwerpen.be}

\begin{abstract}
Domain-specific languages (DSLs) play a crucial role in resolving internal dependencies across enterprises and boosts their upfront business management processes. Yet, a lot of development is needed to build modelling frameworks which support graphical interfaces (canvas, pallettes etc.), hierarchical structures and easy implementation to shorten the gap for novice users. In this paper, a DSL namely, Bmod is introduced, which can be used to model evacuation scenarios. The language is built using Eclipse Modelling Framework (EMF) and Eclipse Graphical Modelling Framework (GMF). Furthermore, a comparison is also shown between Eclipse EMF/GMF and other modelling tools such as AToMPM, metaDepth, Sirius etc with respect to expressiveness, learning curve and performance.
\end{abstract}


\end{frontmatter}



\section{Introduction}
\label{intro}

Contemporary business enterprises and Small and Medium-sized Enterprises (SMEs) face a common challenge in terms of empowering human resources to adapt with the current trends in software modelling and development. Although, the responsibility of building the core software components in an end-product is to be beared by software developers, but domain experts can elude some of the complexities in the modelling process. Meta-modelling and particularly, domain-specific modelling have gradually reduced the cognitive gap between application developers and engineers. Firstly, the huge burden to translate application code into human readable requirements adds a massive delay in the overall engineering process. Secondly, it is quite challenging to design and develop DSLs that translates business/application requirements into their intended software requirements. As such, a DSL is a solution which can allow domain experts to build meta-models as exportable models and later on, software developers can implement the software requirements on top of these models. Therefore, a key requirement in software-driven industries is to build generalized modelling tools and help domain experts accompany the modelling process to yield domain models as per their business requirements. 

\par In this paper, an overview of the Eclipse Modelling Framework (EMF) and Graphical Modelling Framework (GMF) is presented. The paper contemplates on the key features of these modelling frameworks and captures the essence for employing these tools. Furthermore, a DSL namely Bmod [see \ref{description}], which targets to represent evacuation scenarios is built from scratch using Eclipse EMF/GMF tools. The paper provides a walkthrough of the entire language engineering process remarking on some of the key features of EMF/GMF. A comparison is also drawn between EMF/GMF with contemporary modelling tools like AToMPM, metaDepth,Sirius in terms of performance and usability.

\section{Related Work}
\label{litreview}

DSL engineering has become a pervasive task across industries. Recent literature suggests that building graphical tools with user-friendly UI is also a growing demand. In this section, we will discuss the existing literature and some traditional and recent contributions made in extending the paradigm of domain-specific modelling.

\begin{figure}[ht!]
    \centering
    \includegraphics[scale=0.6]{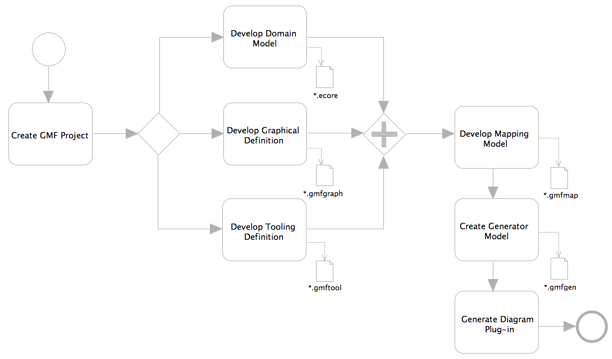}
    \caption{Workflow of Eclipse EMF/GMF}
    \label{fig:overview}
\end{figure}

In \cite{gronback2009eclipse}, Gronback introduced the fundamental aspects to domain-specific modelling, using Eclipse modelling frameworks such as EMF and GMF. To summarize, his work depicted the chronology of developing DSLs [Fig. \ref{fig:overview}] and further illustrated the use of DSLs to accurately deliver domain-specific semantics. Furthermore, Gronback explained model-to-model transformations explicitly using model mappings and refactoring, but did not cover ATL transformations. 

\par With the recent development of powerful frameworks and IDEs, DSLs have been largely democratized across industries. For example, in \cite{hajji2019onto2db}, a language was designed which captured ontological relations and created a conceptual model of relational databases from such ontologies. In \cite{segura2019extremo}, the authors introduced a plugin namely extremo to assist meta-modelling and modelling processes based on extracting embedded information from heterogeneous sources. In \cite{kolovos2010taming}, Kolovos et. al demonstrated EuGENia, a tool to profile a domain model and generate the rest based on model and in-place transformations. However, this work seems obsolete now, since Eclipse IDE provides plugins and wizards to automate this process.
\par In terms of code generation, a comparative analysis of Microsoft DSL tools and Eclipse EMF tools is conducted in \cite{pelechano2006building}. The empirical results suggests Eclipse EMF tools to be more convenient and preferable than MS DSL tools. Especially, this paper identified that amateur users preferred MOFScript language as a template language to build code generators rather than DSL. Furthermore, In \cite{kelly2004comparison}, a comparison between Eclipse EMF and MetaEdit+ was demonstrated by implementing a simple logic gate simulation language with respect to performance only. Quite interestingly, the paper concluded MetaEdit+ to be equally powerful to Eclipse EMF in terms of delivering performance. 

{}

\section{Modelling with Eclipse EMF/GMF}
\label{modelling}

The Eclipse IDE provides modelling frameworks namely Eclipse Modelling Framework (EMF) and Graphical Modelling Framework (GMF) to customize model editors. These editor tools are installed as wizards and plugins, hence allowing users to integrate customized editors and further build or extend meta-models/models of a particular DSL. In this section, a quick overview of the Eclipse EMF platform is provided, which contemplates on the basic prerequisites before building a DSL. In particular, some of the features of EMF/GMF frameworks are covered and a usecase pertaining to evacuation scenarios is presented.

\par The preliminary task to designing any DSL is to design and formulate the abstract syntax of the desired language. For this step, we essentially build the meta-model of our language denoting the features of the elements in our language. For example, in Bmod language, we declare the atomic elements such as floors,rooms,cells,people etc. along with their corresponding attributes and operations. Typically, this is denoted in the language of 'Class Diagrams' in most modelling tools. However, Eclipse EMF/GMF describes the meta-model in Ecore.

\subsection{Ecore meta-modelling language}
\label{Ecore}

The Eclipse Modeling Framework (EMF) includes a meta-model (Ecore) for describing models and runtime support for the models including change notification, persistence support and a very efficient reflective API for manipulating EMF objects generically. 

\begin{figure}[ht!]
\begin{subfigure}{.5\textwidth}
    \centering
    \includegraphics[scale=0.6]{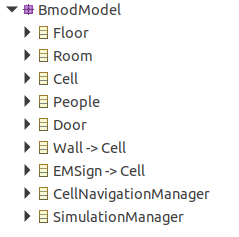}
    \caption{Abstract Syntax of Bmod}
    \label{fig:abstractsyntax}
\end{subfigure}
\begin{subfigure}{.5\textwidth}
    \centering
    \includegraphics[scale=0.58]{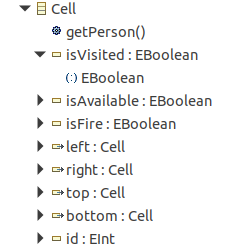}
    \caption{Attributes, Operations and References}
    \label{fig:attributes}
\end{subfigure}{}
\caption{Bmod meta-model in Ecore}
\label{fig:bmod_ecore}
\end{figure}{}

As shown in Fig.\ref{fig:bmod_ecore}, Ecore meta-models depict a tree-based hierarchy structure to denote the elements of the language. The meta-model is essentially constituted of EClasses which further holds instances of type EAttributes, EOperations and EReferences. One of the features of Eclipse EMF supports autonomous transformations of Ecore models into class diagrams [see \ref{viewmodels}]. These are typically stored as separate meta-models known as Ecore Diagrams. Additionally, Eclipse EMF supports storing multiple view models of the abstract syntax (either as class diagrams or tables), which is a handy tool to trace model versions and incremental changes. The above mentioned operations are click-based and does not require manual code.

\par As discussed above, the Eclipse EMF platform provides two modelling formalism namely the Ecore model and the Ecore Diagram model to describe the abstract syntax of our DSL. But the selection of either of these formalism is subject to different usecases and depends on the requirements of the language. As such, the Ecore meta-model provides more expressiveness in terms of properties and features.  The Ecore meta-model allows distinctively to describe attribute types, their upper and lower bounds, default values and accessibility.Hence if the language is required to be acutely property-driven, the Ecore meta-model is preferable. On the other hand, the Ecore Diagram meta-model provides ways to express relationships and containments between different elements of the language. As a result, the Ecore meta-model dominates over Ecore diagram meta-models in terms of expressiveness.

\subsection{Code generation with Genmodel}
\label{genmodel}
The succeeding step after modelling a language is to generate program code which can essentially capture and represent the semantics of the desired language. In Eclipse EMF, this is provided by the Genmodel wizard, which autonomously generates java code from an Ecore meta-model. The Genmodel wizard allows users to generate code for the domain model, the editor and test suites. These are discussed below: 
\subsubsection{Generating Model Code}
\label{modelcode}
The Genmodel wizard essentially generates java classes for each element in the Ecore meta-model. These java classes encapsulates the attributes and the operations, which define the operational semantics of the desired language. Software developers can then extend the generated code to define the operational semantics on top of these java classes, hence preserving the business requirements. In other words, domain experts yield the domain models on top of which operations are executed. And, software developers augment the operational semantics to these entities and describe how those operations are executed by means of code.
\begin{minted}{Java}

package BmodModel;
import org.eclipse.emf.ecore.EObject;
public interface Cell extends EObject {
	boolean isIsVisited();
	void setIsFire(boolean value);
	Cell getLeft();
	void setLeft(Cell value);
	{
	//To write code
	}
	Cell getRight();
	........
	........
	void getPerson();
} // Cell
\end{minted}

The above snippet code gives an overview of the generated java code. Essentially, the code generator adds getter and setter functions for each attribute, along with providing inline documentation of the generated functions.

\subsubsection{Generating Model Editor Code}
\label{modeleditorcode}

After generating all the model artifacts using the Genmodel, the succeeding step is to build the model editor as a plugin. This can be achieved by simply running another eclipse instance with the auto-generated build configuration for the editor bundle. Users can then select the generated Bmod model editor as a seperate wizard to create/extend samples models of the Bmod language.

\par The Eclipse GMF provides a secondary package EMF forms which allows to build and customize view models for Ecore models. A view model essentially models the graphical UI, which is the underlying interface used to modify model attributes and create/delete associations. The EMF forms package additionally automates the UI modelling process by providing generic layouts for each element, such that even non-experts can quickly learn to customize these layouts. The view models are also embedded within the model editor. Additionally, the generated model editor provides UI implementation on top of these customized view models. Hence, the basic CRUD operations and manipulation of sample models becomes more convenient.

\begin{figure}[ht!]
\begin{subfigure}{\textwidth}
    \centering
    \includegraphics[scale=0.3]{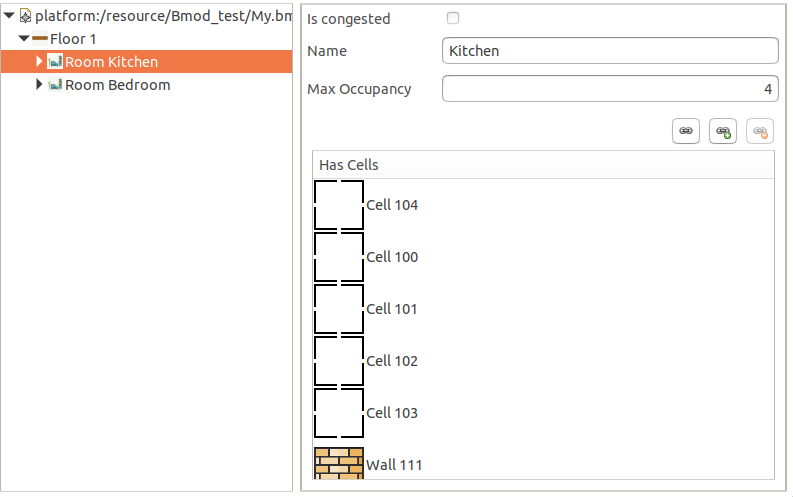}
    \caption{View model for element Room}
    \label{fig:viewmodel1}
\end{subfigure}
\begin{subfigure}{\textwidth}
    \centering
    \includegraphics[scale=0.3]{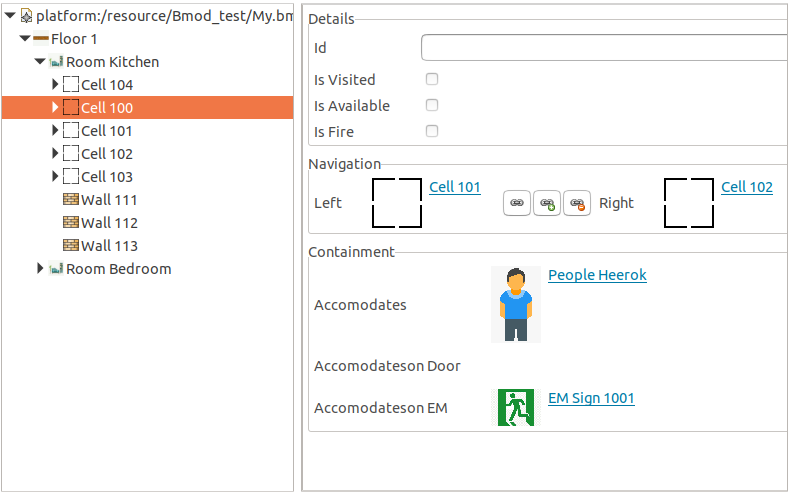}
    \caption{View model for element Cell}
    \label{fig:viewmodel2}
\end{subfigure}{}
\caption{Customized view models}
\label{fig:viewmodel}
\end{figure}{}

\subsubsection{Generating Diagram Editor Code}
\label{diaeditorcode}
The Eclipse EMF/GMF tool provide plugin support to build customized graphical and model editors. The model editor can be comfortably generated using the GMF dashboard. After building the Ecore meta-model of the DSL, users can derive the generator model using the dashboard.

\begin{figure}[ht!]
    \centering
    \includegraphics[scale=0.5]{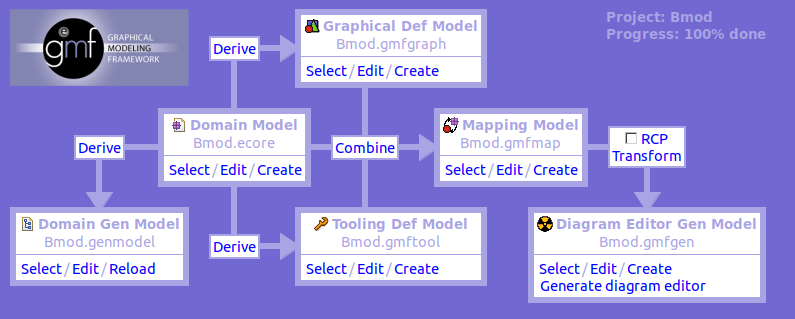}
    \caption{Eclipse GMF Dashboard}
    \label{fig:gmfdashboard}
\end{figure}{}
\par Fig. \ref{fig:gmfdashboard} illustrates the GMF dashboard. As observed in the figure, the only pre-requisite artifact required to generate diagram editors and model editors is the domain model. The domain model is built employing Ecore meta-modelling language, which can be easily built by domain experts. Since, the succeeding steps are derived solely from the domain model, the entire code generation process becomes autonomous relieving domain-experts from acquiring additional skills and workload. Based on the generated editors, software developers can consequently model the software requirements and add necessary implementation code to achieve the desired semantics of the language.

\section{Implementation of Bmod using EMF/GMF}
\label{bmod}

In this section, a chronological summary of the entire project is discussed including anecdotal remarks and technical difficulties faced while modelling the Bmod language.

\subsection{Defining Abstract Visual Syntax}
\label{avs}

\begin{figure}[ht!]
    \centering
    \includegraphics[scale=0.45]{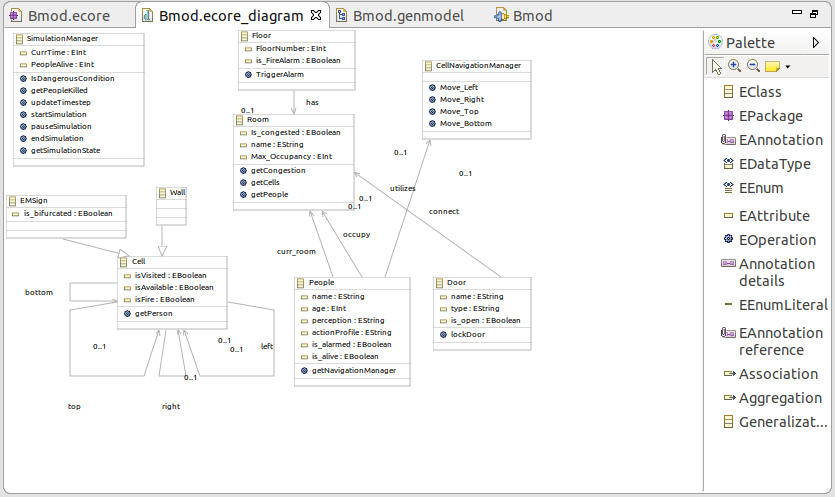}
    \caption{Abstract Visual Syntax of Bmod}
    \label{fig:avs_bmod}
\end{figure}

The abstract visual syntax defines the declarative elements of a language. As such for our desired language Bmod, we have used the Ecore Diagram Editor to define the language elements. As shown in Fig.\ref{fig:bmod_ecore} and Fig.\ref{fig:avs_bmod}, the ecore model depicts the atomic elements of our DSL such as Floors, Rooms, Cells, Door People, Emergency Signs and Walls. 

{}

\subsection{Defining Concrete Visual Syntax}
\label{cvs}
The concrete visual syntax defines the visual representation of each element of our DSL. The Genmodel wizard generates the edit code that automatically maps every icons for every declared element including references, associations and classes. The generated code is stored as a separate bundle in the Eclipse IDE and users can modify icons directly from the directory. As such, for defining the concrete syntax for Bmod, images were exported as gifs and later on replaced with the existing icons. However, in terms of expressiveness, Eclipse GMF does provide extensive graphical definition tools which support complex polygon structures, floating texts and embedded shapes and figures. Alternatively, an easy and comfortable approach is to derive this functionality from the Genmodel wizard, hence refraining domain-experts to indulge with these sophisticated tools. 

\begin{figure}[ht!]
    \centering
    \includegraphics[scale=0.4]{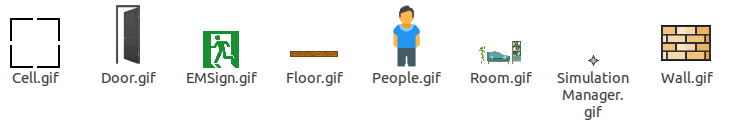}
    \caption{Overview of Concrete Visual Syntax of Bmod}
    \label{fig:cvs}
\end{figure}{}

\subsection{Creating Sample Models of Bmod}
\label{samplemodels}

Employing the generated model editors and diagram plugins, sample models can be built from scratch. These operations are mostly click-and-drag operations and does not require any manual coding. Fig. \ref{fig:viewmodel} illustrates some samples models generated using the model editor wizard.

\begin{figure}[ht!]
    \centering
    \includegraphics[scale=0.4]{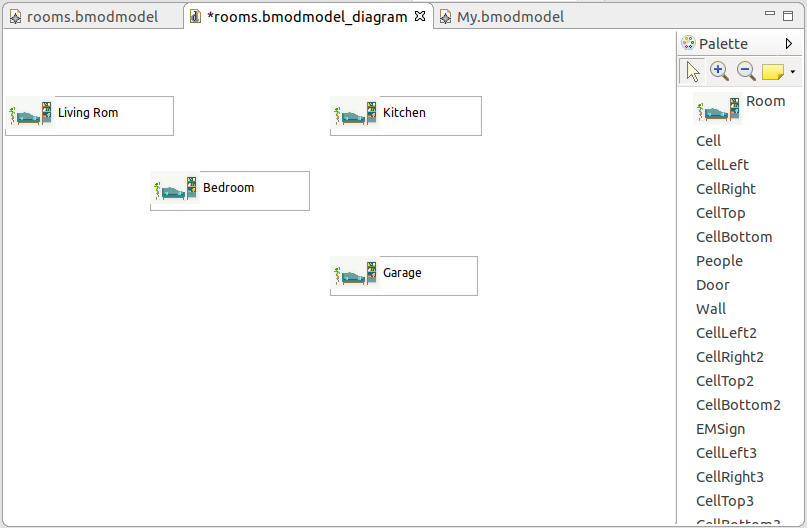}
    \caption{Bmod Diagram model}
    \label{fig:diagram_editor}
\end{figure}{}

Fig. \ref{fig:diagram_editor} illustrates a sample model built using the generated diagram editor. In terms of expressiveness, we observe that the model fails to capture the essence of containments and associations. Due to limited time and the inherent complexity, especially while defining the diagram definition, the model does not represent references. Although, it is difficult to represent links between elements as a novice user, associations can indeed be represented in model diagrams. However, examples and online resources positively indicate that it is difficult to represent containment relationships in \cite{biermann2006graphical}.

\section{Comparative Analysis}
\label{compare}
In this section, a comparison of Eclipse EMF/GMF tool is presented with some of the state-of-the-art modelling tools. As such, the comparisons will be strictly inclined towards Eclipse EMF/GMF and AToMPM \footnote{Other MDE tools are neglected due to lack of practical experience in these tools}.

\begin{table}[ht!]
\footnotesize
\caption{Comparison of different MDE tools}
\begin{tabular}{@{}cccc@{}}
\toprule
\multirow{2}{*}{MDE Tool} & \multirow{2}{*}{\begin{tabular}[c]{@{}c@{}}Domain \\ Model\end{tabular}} & \multirow{2}{*}{\begin{tabular}[c]{@{}c@{}}Code\\ Generation\end{tabular}} & \multirow{2}{*}{\begin{tabular}[c]{@{}c@{}}Model \\ Transformation\end{tabular}} \\
                          &                                                                          &                                                                            &                                                                                  \\ \midrule
Eclipse EMF/GMF           & Ecore models                                                             & \begin{tabular}[c]{@{}c@{}}Genmodel\\ (fully autonomous)\end{tabular}      & ATL                                                                              \\ \\
AToMPM                    & Class Diagrams                                                           & Manual                                                                     & Rule-based, MoTIF                                                                \\ \\ 
Eclipse Sirius            & Ecore (Diagram)                                                          & Genmodel                                                                   & Acceleo/ ATL                                                                     \\ \\ 
Eclipse Graphiti          & Ecore (Diagram)                                                          & Underlying EMF/GMF                                                         & Not supported                                                                    \\\\ 
XText                     & Textual                                                                  & XTend                                                                      & \begin{tabular}[c]{@{}c@{}}ATL (via exporting \\ models)\end{tabular}            \\  \\
metaDepth                 & Textual                                                                  & Not Supported                                                              & ETL/Epsilon                                                                      \\ \bottomrule
\end{tabular}
\label{comparison:MDEtools}
\end{table}

Table \ref{comparison:MDEtools} gives an overview of the underlying meta-models, code generators and model transformation languages for the aforementioned tools. With respect to code generation, Eclipse EMF/GMF dominates over other tools since the primary focus of EMF/GMF is to reduce the effort in code generation. The Genmodel wizard is much more convenient to use as compared to AToMPM. AToMPM currently requires manual code intervention, specifically transforming meta-models into sourceTree models before generating python code. This would require some trivial effort, if any, to further customize the generated code such as adding annotations, inline comments and additional implementation code.

\par In terms of model transformation, AToMPM is much more convenient since it support visual rule-based transformations. Firstly, Eclipse EMF/GMF does not support endogenous model-to-model transformations whereas AToMPM supports both endogenous and exogenous transformation. Secondly, the transformation language used in Eclipse EMF/GMF is ATLAS (ATL), which is purely text-based and requires some additional effort to learn before applying practically. On the other hand, AToMPM delivers an easy and much comprehensible visual editor to create pattern/rule-based transformations. Additionally, the underlying MoTif scheduling language helps to incorporate the operational semantics of a DSL.

\begin{table}[ht!]
\caption{Comparison of different model features}
\begin{tabular}{@{}ccccc@{}}
\toprule
\multirow{2}{*}{MDE Tool} & \multicolumn{4}{c}{Model Features}                     \\ \cmidrule(l){2-5} 
                          & Expresiveness & Navigability & Hierarchy & Refactoring \\ \cmidrule(r){1-5}
Eclipse EMF/GMF           & High          & High         &   $\surd$        &     $\surd$        \\
AToMPM                    & High          & Low         & $\surd$           &     $\times$        \\
Eclipse Sirius            & High          & High         &  $\surd$         &      $\surd$       \\
Eclipse Graphiti          & High          & Low          &   $\times$        &     $\surd$        \\
XText                     & Low           & Low          &     $\surd$      &      $\surd$       \\
metaDepth                 & Low           & Low          &    $\surd$       &   $\surd$          \\ \bottomrule
\end{tabular}
\label{comparison:modelfeatures}
\end{table}

\par Table \ref{comparison:modelfeatures} compares some of the features of sample models with respect to visualization and delivery of the intended semantics. In terms of expressiveness and representation, although Ecore models dominate as opposed to traditional class diagrams, the visual representation in AToMPM is much convincing than that of Eclipse EMF/GMF. However in terms of data modelling, Eclipse EMF/GMF is slightly better, as it provides tree-based hierarchical structures along with easy create/delete/modify functionalities. Additionally, customized model editors along with automated UI makes it more convenient to navigate and modify the sample models.

\section{Conclusion}
\label{conclusion}

In this paper, a brief summary of domain-specific modelling usecase is exemplified by implementing a DSL for modelling evacuation scenarios. The paper discussed key features of Eclipse EMF/GMF such as automated code generator and GMF dashboard to ease the efforts in domain-specific modelling. Furthermore, the paper presented anecdotal remarks on the tool and compared its performance and usability with other contemporary modelling tools. Clearly, ATomPM is a much more user-friendly tool as compared to Eclipse EMG/GMF to model and integrate operational semantics of a DSL, but, it lacks support for automated code generation. Nevertheless, Eclipse EMF/GMF is a powerful modelling framework focusing on code generation and customized model/graphical editors, yet it is considerable to augment more features.





\bibliography{cite}
\bibliographystyle{elsarticle-harv}
\appendix

\section{Description of Bmod DSL}
\label{description}

This section describes the set of elements that constitute the Bmod language and their corresponding semantics. The primary use case of Bmod language is to build models that create evacuation scenarios and operate on these models to analyse behaviour of participants, detect alarming events and perform safety analysis of floor plans.

\begin{itemize}
    
\item A Floor is one of the elements of Bmod language, which is hierarchically above every other element. An instance of floor denotes a regular floor of a building.

\item A Room denotes a finite space, which encapsulates a set of cells, doors, people and an emergency situation \textit{(In this case, we assume its only fire)}.

\item People denote instances of humans that are present in a room. People have a perception and an action profile while influence their navigation during the evacuation scenario. People are either alive or killed by fire.

\item A Door denotes a gateway from one room to another. It encapsulates a hypothetical path between one source room and a destination room. A door is exists as either locked or open. An open door allows people to transport from one room to another.

\item A Cell denotes a navigational element, which is used by people to escape during the evacuation event. Every cell is connected to one cell in all the four directions. A cell accommodates people, emergency signs, doors and fire. Essentially, a set of cells constitutes a room.

\item A Wall denotes an obstruction. A wall obstructs people and fire to transport to other cells. Walls are essentially derived from cells, but they are not connected to any other cells in all four directions.

\item A EMSign denotes an emergency sign that guides people to the exit. A set of emSign denotes a evacuation path, which can be used by the people to escape.

\item CellNavigationManager denotes an abstract class which helps to obtain navigational information. This will be required while simulating an evacuation scenario, where people would move from one cell to another.

\item SimulationManager denotes another abstract class that represents the simulation state of the scenario. This class contains variables denoting time, validation attributes and operations to pause/resume simulation.
\end{itemize}

\section{Representations of Ecore models}
\label{viewmodels}

\begin{figure}[ht!]
\begin{subfigure}{\textwidth}
    \centering
    \includegraphics[scale=0.5]{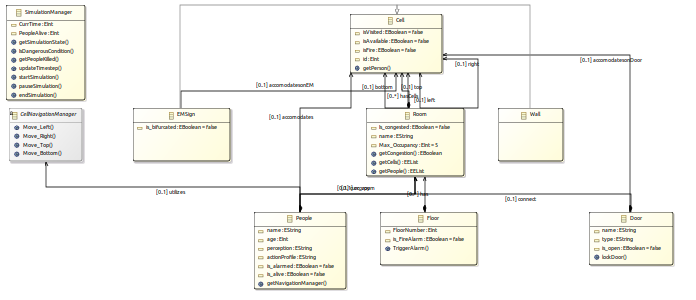}
    \caption{Class diagram representation}
    \label{fig:CDview}
\end{subfigure}
\vspace{1em}
\begin{subfigure}{\textwidth}
    \centering
    \includegraphics[scale=0.3]{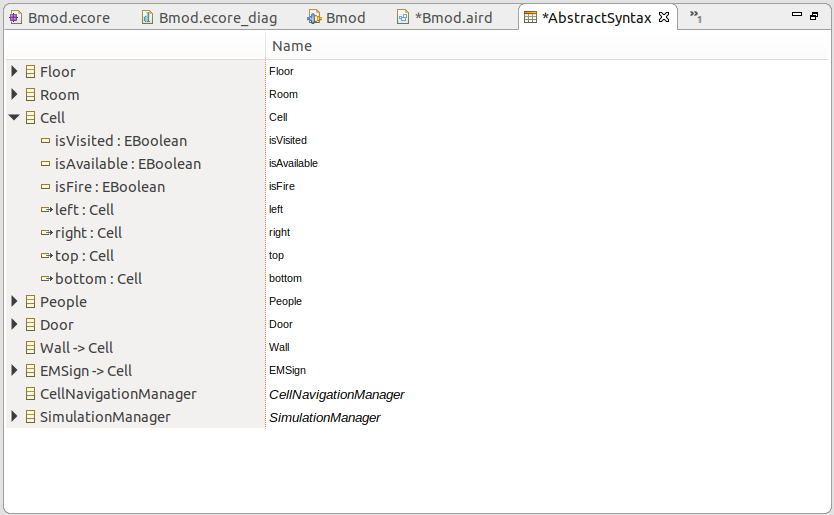}
    \caption{Spreadsheet representation}
    \label{fig:spreadview}
\end{subfigure}{}
\caption{Different view representations}
\label{fig:view_models}
\end{figure}

 The Eclipse EMF tool allows users to create multiple view representations of the base Ecore model.  As shown in Fig. \ref{fig:view_models}, the Ecore model can be viewed either as a class diagram or as a spreadsheet. Additionally, multiple instances of the Ecore model can be stored. This is a useful feature to trace model updates and versions throughout the SDLC lifecycle.

{}

\end{document}